\documentclass[preprint]{aastex631}

\usepackage{soul}
\usepackage{hyperref}
\usepackage{xcolor}
\usepackage{amsmath}

\newcommand{\rev}[1]{#1}

\received{May 8, 2025}
\revised{January 5, 2026}

\submitjournal{ApJ}

\begin{document}
\title{\rev{Time-Dependence of Subsurface Solar Convection} Using the Time-Distance Deep-Focus Method}

\correspondingauthor{John T. Stefan}

\author[0000-0002-5519-8291]{John T. Stefan}
\affiliation{New Jersey Institute of Technology, University Heights, Newark, NJ 07102, USA}
\email{john.stefan@njit.edu}

\author[0000-0003-0364-4883]{Alexander G. Kosovichev}
\affiliation{New Jersey Institute of Technology, University Heights, Newark, NJ 07102, USA}
\affiliation{NASA Ames Research Center, Moffett Field, Mountain View, CA 94043, USA}

\author[0000-0002-2671-8796]{Gustavo Guerreo}
\affiliation{Universidade Federal de Minas Gerais, Av. Antonio Carlos 6627, Belo Horizonte, MG 31270-901, Brazil}

\author[0000-0001-7483-3257]{Andrey M. Stejko}
\affiliation{New Jersey Institute of Technology, University Heights, Newark, NJ 07102, USA}

\begin{abstract}
We re-examine the deep-focus methodology of time-distance helioseismology previously used to estimate the power spectrum of the solar convection at a depth of about 30 Mm, which was found to be significantly weaker than predicted by theory and simulations. The Global Acoustic, Linearized Euler (GALE) and Eulerian Lagrangian (EULAG) codes are used to generate ground-truth simulations to evaluate the accuracy of the inferred convective power spectrum. This validation process shows that the power spectrum \rev{derived using the time-distance methodology} diverges significantly from ground truth beyond spatial scales corresponding to the spherical harmonic degree $\ell=15$--$30$ because of the limited resolution of helioseismic measurements at that depth. However, the power estimated at larger spatial scales ($\ell<15$) is sufficiently accurate. We then apply the methodology to solar data \rev{selected from throughout Solar Cycle 24 and find some evidence that the magnitude of the convective power changes throughout the Cycle. An average of the convective power across the Solar Cycle reveals a spectrum that is qualitatively similar to previous estimates, though about half an order of magnitude greater. The disagreement between observations of solar convection and the magnitudes predicted by simulations persists.}
\end{abstract}


\section{Introduction}\label{sec:intro}

It is well-known that convection within the Sun operates on a wide range of spatial scales and amplitudes. This is most clearly seen in the photosphere, where convective motions can be discerned from direct Doppler velocity observations and from local correlation tracking and other methods applied to optical observations. Such methods reveal several distinct scales, including granules with spatial scales of 1\,--\,2\,Mm, supergranules with spatial scales of about 30\,Mm \citep{Leighton1962}, and giant cells with spatial scales of several hundred Mm \citep[e.g.,][]{Hathaway2013,Hathaway2021}. While giant banana-shaped cells are clearly observed in simulations of global convection \citep{Miesch2008}, the observational evidence for their existence still remains ambiguous. Still, simulations can reproduce small-scale solar surface convection remarkably well \citep[see][for a full review]{Nordlund2009}.



Recent radiative hydrodynamic simulations of the near-surface convection in the presence of rotation revealed a significant role of the hydrogen and helium ionization zones \citep{Kitiashvili2023, Kitiashvili2025}. In particular, the turbulent spectra show an increase in scale with depth and a qualitative change in convective patterns below 7 Mm (near the bottom of the so-called leptocline), suggesting changes in turbulent diffusivity and energy exchange across scales. This is in stark contrast to global simulations, which are performed in an anelastic approximation and do not include the near-surface layers. Characteristic flows in these simulations are strongly influenced by the Rossby number \citep{Guerrero2013,Featherstone2015}. In particular, the differential rotation profile transitions from a fast equator, slow pole state (solar-like) at $R_o\ll$1 to a slow equator, fast pole state (anti-solar) at $R_o>1$. To remain in the Rossby number regime as the Sun, global models of convection either increase the rotation rate \citep{Augustson2015} or decrease the luminosity \citep{Hotta2015}.

Meanwhile, measurements of deep convection in the Sun also have their own challenges, and these measurements rely on several different helioseismic techniques since direct investigation is not possible. The helioseismic measurements have been focused on estimating the convective power spectrum at a depth of about 30 Mm ($\sim 0.96\,R_\odot$), where these measurements can be compared with the global Sun simulations.  These techniques include inversions of the subsurface flows from acoustic wave travel times \citep{Getling2022}, direct analysis of the acoustic wave travel times \citep{Hanasoge2012}, or inference of the subsurface flows from the Doppler effect observed by the ring-diagram local helioseismology method \citep{Greer2015,Proxauf2021}. Each of these methods is sensitive to different ranges of spatial scale, though crucially they have overlapping sensitivity beyond $\ell\approx10$. This should, in theory, allow for a more constrained measurement of the power spectrum in this range. However, recently revised estimates of the power spectra obtained from the ring-diagram analysis \citep{Greer2015} and from direct analysis of the acoustic wave travel times \citep{Hanasoge2012} still differ by several orders of magnitude \citep{Proxauf2021,Birch2024}.

In the context of the disparate convective power spectrum measurements as well as the fine-tuning problem facing global simulations, the \textbf{Convective Conundrum} is succinctly stated by \citep{Omara2016}: "The convective velocities required to transport the solar luminosity in global models of solar convection appear to be systematically larger than those required to maintain the solar differential rotation and those inferred from solar observations." There are, then, two facets of the convective conundrum: the incompatibility of the magnitude of modeled convective velocities with the formation of solar-like differential rotation, and the currently large discrepancy in the solar convective power spectrum as measured by different methods. There has recently been some progress on the modeling aspect of the convective conundrum, where less finely-tuned simulations have been able to produce both solar-like differential rotation and convective flows with reasonable amplitudes \citep{Hotta2021,Noraz2025}.

It is our goal in this work to investigate the uncertainties associated with the helioseismic measurements of solar convective power, specifically from the deep-focus time-distance helioseismology technique employed by \citet{Hanasoge2012}. To that end, we perform validation of the procedure using acoustic simulations generated by the GALE code \citep{Stejko2021} using a background flow field derived from EULAG simulations of global convection \citep{Guerrero2022}. This flow field serves as a "ground truth" against which the spectrum measured from the acoustic wave travel times can be compared. In Section \ref{sec:df_method}, we describe the time-distance helioseismology methodology used to obtain the travel times associated with longitudinal flows; for reproducibility, we provide the exact parameters of our measurement geometry. Section \ref{sec:cal_method} summarizes the analysis from \citet{Hanasoge2012} that enables the power spectrum computation, and Section \ref{sec:obs_method} briefly describes how we handle solar observations. We then present the results of our calibration and validation procedures in Section \ref{sec:cal_results}, followed by the new estimate for the solar convective power spectrum in Section \ref{sec:obs_results}.

\section{Methods} \label{sec:methods}

\subsection{Principles of Deep-Focus Helioseismology} \label{sec:df_method}

In general, the one-way travel time of an acoustic wave is perturbed by some amount $\delta\tau$ as it travels through variations in the background sound speed and flows, given by \citep{Kosovichev1997}
\[
\delta \tau = -\int_\Gamma \left[\dfrac{\mathbf{n}\cdot\mathbf{u}}{c^2}+\dfrac{\delta c}{c^2} \right]ds,
\]
where $\mathbf{n}=\mathbf{k}/|\mathbf{k}|$ is the direction of wave propagation at point $s$ along the unperturbed ray path $\Gamma$, $\mathbf{u}$ is the background flow field, $c$ is the unperturbed sound speed, and $\delta c$ is the perturbation of the sound speed. The flow signal can be isolated by taking the difference of the travel time deviations from waves traveling in opposite directions, i.e.
\begin{equation}\label{eqn:dtflow}
\delta\tau_{\text{flow}} = \dfrac{1}{2}\left(\delta\tau^{\text{(forward)}}-\delta\tau^{\text{(backward)}}\right) = -\int_\Gamma \dfrac{\mathbf{n}\cdot\mathbf{u}}{c^2}ds .
\end{equation}

Several different measurement geometries can be used to obtain travel time perturbations---each designed to improve the signal-to-noise ratio (SNR) in the perturbations associated with particular quantities at some depth---and in line with the methodology described by \citet{Hanasoge2012}, we use a deep focusing method. This method averages the travel-time deviations obtained from waves that propagate to a variety of depths, but which all intersect at a desired spatial location and depth, which we refer to as the focal point. We determine the particular ray paths and corresponding measurement geometry by iterating over starting points away from the submerged focal point in increments of 0.6 heliographic degrees, up to a distance of 30$^\circ$. Given these starting locations, the equation for the angular half-travel distance of an acoustic wave with turning point $R^\star$ is incrementally solved over radius for all possible turning points; this equation is given by
\begin{equation}\label{eqn:halfdist}
\dfrac{1}{2}\Delta(R^\star) = \int_{R^\star}^{R_{\odot}}\dfrac{c}{w_\text{ph}r^2}\dfrac{dr}{\sqrt{1-\left(\dfrac{c}{w_\text{ph}r}\right)^2}},
\end{equation}
where $\Delta(R^\star)$ is the full angular travel distance of an acoustic wave with turning point $R^\star$, $c$ is the local sound speed, and $w_\text{ph}=c(R^\star)/R^\star$ is the angular phase speed of the wave at the turning point in units of rad\,s$^{-1}$\,$\ell^{-1}$. The ray path is symmetric about the turning point, and the full travel distance of the acoustic wave is simply twice Equation \ref{eqn:halfdist}.

\begin{figure}
    \centering
    \includegraphics[width=0.5\linewidth]{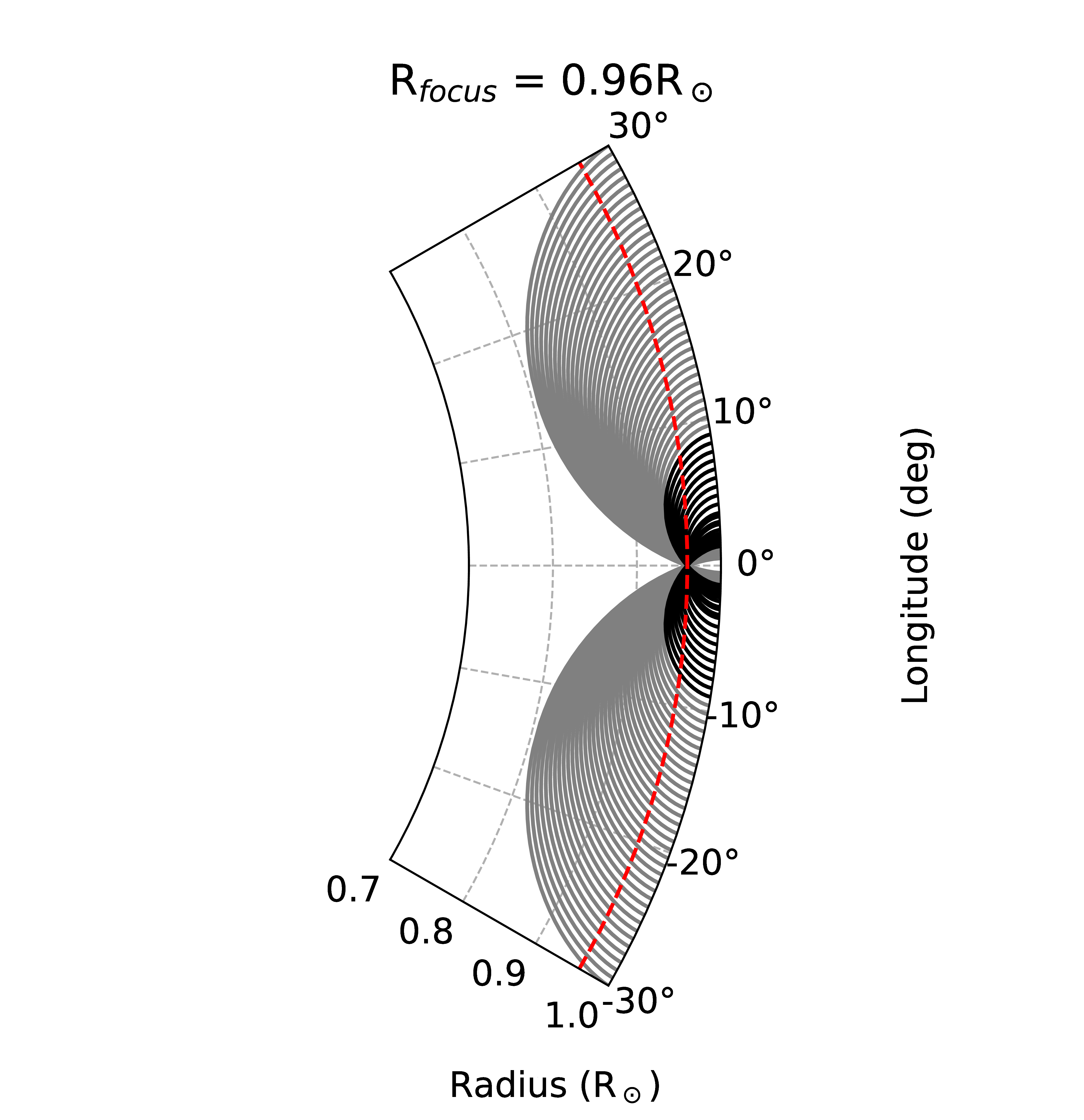}
    \caption{Acoustic waves that intersect the focal point at radius 0.96R$_\odot$. We discard wave paths that intersect the focal point at an angle greater than 45$^\circ$ (\textit{gray lines}) and keep only those that intersect at an angle less than 45$^\circ$ (\textit{black lines}). The focal depth is indicated by the \textit{red dashed} line.}
    \label{fig:measurement}
\end{figure}

We identify the ideal ray path for each starting distance from the above equation as the one which minimizes the horizontal distance from the focal point when the wave reaches the focal depth $R_\text{f}$, and the resulting ray paths are displayed in Figure \ref{fig:measurement}. The ray path selection is further refined according to two criteria; first, we find that there is no appropriate measurement geometry for starting locations less than 3$^\circ$ as these rays do not pass through the focal point. Second, the travel time perturbation as described in Equation \ref{eqn:dtflow} is dependent on the angle at which the acoustic wave intersects the background flow. There is presumably some critical angle at which the perturbation is more sensitive to the radial component of the flow rather than the horizontal component. Determining this critical angle and the associated effect on SNR are beyond the scope of this work, and we choose $\theta_i=\arctan(k_r/k_h)=45^\circ$ as a threshold for the allowable angle of incidence at the focal point. The ray paths that satisfy the condition are displayed in Figure \ref{fig:measurement}, and the remaining ray paths are displayed in gray.

We can use the identified acoustic waves to determine the distances at which we should cross-correlate the surface velocity---we specifically correlate the radial component of the surface velocity for simulation data and the line-of-sight (LOS) velocity for observational data. The start distance is given by the initial increment of distance as described above, and the end distance by the difference between the start distance and total travel distance derived from Equation \ref{eqn:halfdist}. Instead of computing the point-to-point cross-correlation, we average the velocity signal within arcs $\pm$45$^\circ$ of the East and West directions, respectively, and cross-correlate these averaged signals over a time interval $T$ that is chosen to accumulate a sufficient signal-to-noise ratio (SNR).

Before taking such measurements, however, we apply a Gaussian phase speed filter to the velocity data to further improve our SNR. The filter is centered at $w_\text{ph}$ for the particular acoustic wave, with width $\sigma=0.05w_\text{ph}$. For each identified ray path, the surface velocity is phase-speed filtered, and the cross-correlation procedure is applied. In general, the resulting cross-correlation functions are best described by a Gabor wavelet \citep{Kosovichev1997,Nigam2007} that depends on several factors, including the phase travel time $\tau_\text{ph}$. We average the cross-correlations over the identified ray paths after first shifting them according to the difference in unperturbed phase travel time relative to the mean. The unperturbed travel time is computed by
\[
\tau_\text{ph}(R^\star) = 2\int_{R^\star}^{R_\odot}\dfrac{1}{c}\dfrac{dr}{\sqrt{1-\left(\dfrac{c}{w_\text{ph}r}\right)^2}}.
\]
We then fit a Gabor wavelet to the computed cross-correlations to obtain the phase travel time for both the forward (East-West) and backward (West-East) directions, and the flow travel time perturbation is obtained from the difference of these two times as in Equation \ref{eqn:dtflow}. The parameters of our measurement procedure, including the $w_\text{ph}$s and unperturbed travel times, are listed in Table \ref{tab:measurement}.

\begin{table}
    \centering
    \begin{tabular}{||c||c|c|c|c|c|c||}
    \hline
         Index & $R^\star$ ($R_\odot$) & $\Delta$ (deg) & $\tau_\text{ph}$ (mins) & Starting Distance (deg) & $w_\text{ph}$ ($\mu$Hz/$\ell$) & $\theta_i(R_\text{f})$ (deg)\\
         \hline\hline
         1 & 0.9595 & 6.43 & 45.8 & 3.6 & 16.36 & 7.7 \\
\hline
2 & 0.9582 & 6.65 & 46.5 & 4.2 & 16.69 & 13.8 \\
\hline
3 & 0.9562 & 6.98 & 47.3 & 4.8 & 17.20 & 19.5 \\
\hline
4 & 0.9538 & 7.37 & 48.4 & 5.4 & 17.79 & 24.3 \\
\hline
5 & 0.9513 & 7.80 & 49.5 & 6.0 & 18.41 & 28.3 \\
\hline
6 & 0.9485 & 8.27 & 50.6 & 6.6 & 19.08 & 31.8 \\
\hline
7 & 0.9456 & 8.76 & 51.7 & 7.2 & 19.76 & 34.9 \\
\hline
8 & 0.9425 & 9.27 & 52.9 & 7.8 & 20.47 & 37.6 \\
\hline
9 & 0.9394 & 9.79 & 54.1 & 8.4 & 21.19 & 40.1 \\
\hline
10 & 0.9362 & 10.32 & 55.2 & 9.0 & 21.90 & 42.2 \\
\hline
11 & 0.933 & 10.85 & 56.3 & 9.6 & 22.60 & 44.2 \\
\hline
    \end{tabular}
    \caption{Parameters for the measurement procedure used in this work. $\Delta$ is the total angular distance traveled by the particular mode, $\tau_\text{ph}$ is the mode's travel time, $w_\text{ph}$ is the angular phase speed, and $\theta_i(R_\text{f})$ is the angle of incidence at the focal depth.}
    \label{tab:measurement}
\end{table}

\subsection{Calibration Steps for an Upper Bound on the Convective Power Spectrum} \label{sec:cal_method}

Our ultimate goal here is to provide an upper bound on the convective velocities at various characteristic spatial scales. In this work, we follow the methodology outlined by \citet{Hanasoge2012}, which we briefly describe in this section. We wish to connect the interior flow velocity, $\mathbf{v}$, to the forward-backward travel-time difference in Equation \ref{eqn:dtflow}; we denote the travel time shift due to the $i^{\text{th}}$ component of the flow as $\delta\tau^{(i)}$. The general relationship is given by
\begin{equation}\label{eqn:kernel}
    \delta\tau^{(i)} = \int_0^{R_\odot}\int_0^\pi\int_0^{2\pi}\left(\mathbf{K}^{(i)}\cdot\mathbf{v}\right)  \dfrac{r^2 dr}{V}\sin\theta d\phi d\theta ,
\end{equation}
where $\mathbf{K}^{(i)}$ is the kernel that describes the sensitivity of the measurement procedure to the $i^{\text{th}}$ component of the flow, having units s\,(km\,/\,s)$^{-1}$, and $V$ is the total volume of the Sun. In these units, the travel time is given by the volume-averaged response to the background flow. The sensitivity kernel is a vector such that $K^{(i),p}$ is the sensitivity of the measurement of the $i^{\text{th}}$ component of the flow to the $p^{\text{th}}$ component of the actual flow. In other words, the elements of $K^{(i),p}$ where $i\ne p$ correspond to the so-called "leakage" of other components of flow into the desired measurement. While there are many formulations for these kernels---for example, the ray-path \citep{Kosovichev1996} and the Born \citep{Gizon2002} formalisms---the methodology followed here is structured so that a precise formulation of the sensitivity kernels is not necessary.

Let us now move to the spherical harmonic domain, where we can represent flow components in terms of spherical harmonic coefficients (SHCs) computed by
\[
A_{\ell,m} =\int_{0}^{2\pi} \int_0^{\pi}Y_{\ell,m}^\star(\theta,\phi) f(\theta,\phi)\sin\theta d\theta  d\phi,
\]
and the mean square value of the function is
\[
\langle f^2\rangle = \dfrac{1}{4\pi}\int_0^{2\pi}\int_0^\pi f^2\sin\theta d\theta d\phi.
\]
We choose to use the orthonormalized spherical harmonics, such that integrating the product of any two harmonics $(\ell,m)$ and $(\ell^\prime,m^\prime)$ over the sphere yields
\[
\int_0^\pi\int_0^{2\pi}Y_{\ell,m}^\star Y_{\ell^\prime,m^\prime}\sin\theta d\phi d\theta = \delta_{\ell,\ell^\prime}\delta_{m,m^\prime},
\]
where $Y_{\ell,m}^\star$ is the complex conjugate of $Y_{\ell,m}$. The SHCs of the measured travel times \rev{associated with the $\phi$-component of flow}, as expressed in Equation \ref{eqn:kernel}, are then
\[
\delta\tau_{\ell,m}^{(i)} = \int_0^{R_{\odot}}(\mathbf{K}_{\ell,0}^{(i)}\cdot \mathbf{v}_{\ell,m})\dfrac{4\pi r^2 dr}{V}.
\]
Here, we have neglected the $m\ne 0$ components of the sensitivity kernel, as it has been shown that these components contribute very little to the overall sensitivity \citep{Hanasoge2012}. 

Following the methodology of \citet{Hanasoge2012}, we assume that the flows at different depths are independent of each other, such that $v(r)v(r^\prime) = v(r)^2\delta(r-r^\prime)$ with the delta function defined in spherical coordinates as 
\[
\int_0^{R_{\odot}}\delta(r)\dfrac{4\pi r^2 dr}{V} = 1.
\]
It can then be shown that the square of the travel time deviations associated with the longitudinal component of flows satisfies the following inequality,
\begin{equation}\label{eqn:ineq_beg}
 (\delta\tau_{\ell,m}^{(\phi)})^2 > \int_0^{R_\odot}(K_{\ell,0}^{(\phi),\phi})^2 (v_{\ell,m}^\phi)^2\dfrac{4\pi r^2 dr}{V} ,
\end{equation}
with $K^{(\phi),\phi}$ being the measurement sensitivity to the $\phi$-component of flow for the same component of the actual flow. The above inequality can be further weakened by integrating only in some small neighborhood of the target depth,
\begin{equation}\label{eqn:ineq_int}
 (\delta\tau_{\ell,m}^{(\phi)})^2 > \int_{D-\epsilon}^{D+\epsilon}
(K_{\ell,0}^{(\phi),\phi})^2 (v_{\ell,m}^\phi)^2 \dfrac{4\pi r^2 dr}{V}.
\end{equation}

Suppose that the radial and $\ell$ dependence of the kernel can be separated such that $K_{\ell,0}^{(\phi),\phi)} = A_{\ell}f(r)$ in the integration range. Note that this is a slightly different expression for the kernel than the one used in \cite{Hanasoge2012}, though the end result is similar. The above inequality can then be expressed as
\[
 (\delta\tau_{\ell,m}^{(\phi)})^2 > \int_{D-\epsilon}^{D+\epsilon}
A_{\ell}^2f^2(r) (v_{\ell,m}^\phi)^2 \dfrac{r^2 dr}{V} = A_{\ell}^2  (v_{\ell,m}^\phi)^2\int_{D-\epsilon}^{D+\epsilon} f^2(r) \dfrac{4\pi r^2 dr}{V},
\]
where the right-hand equality arises when the bounds of integration are less than the pressure scale height, across which the flow velocities are not expected to vary. Defining
\[
C_\ell^2 = A_\ell^2\int_{D-\epsilon}^{D+\epsilon} f^2(r) \dfrac{4\pi r^2 dr}{V},
\]
we can then express the inequality from Equation \ref{eqn:ineq_int} as
\begin{equation}\label{eqn:ineq}
 (\delta\tau_{\ell,m}^{(\phi)})^2 > C_\ell^2  (v_{\ell,m}^\phi)^2.
\end{equation}
Therefore, if the calibration coefficients $C_\ell$ can be found, then the acoustic wave travel times can provide an upper bound on the convective power spectrum,
\begin{equation}\label{eqn:ineq_mid}
\dfrac{ (\delta\tau_{\ell,m}^{(\phi)})^2}{C_\ell^2} >  (v_{\ell,m}^\phi)^2.
\end{equation}

In a realistic setting, however, it was found that additional factors must be introduced to Equation \ref{eqn:ineq_mid} that account for the signal-to-noise ratio (SNR) as well as the restriction of the integral in Equation \ref{eqn:ineq_int}, where presumably the flows around the focus depth occupy some characteristic range greater than the integration range, for example the mixing length \citep{Proxauf2021}. While in principle we cannot restrict acoustic waves to travel only through a given range of depths so that the inequality in Equation \ref{eqn:ineq_beg} transforms to Equation \ref{eqn:ineq_int}, we can artificially impose radially-localized flows in simulations so that the integration bounds in Equation \ref{eqn:ineq_int} hold. A delta function is best-suited for this purpose, but to avoid numerical artifacts, we treat the source flow as a Gaussian in both the horizontal and radial directions; for the focus depth of $0.96R_\odot$, the radial and horizontal widths are $\sigma_r=1.74$~Mm and $\sigma_h=3\sigma_r$, respectively.

We place 500 of these "delta" flows throughout the simulation domain centered at $r=0.96R_{\odot}$, each with a randomly-chosen location in latitude and longitude. The amplitude of the flows is chosen to be $5\%$ of the sound speed at $r=0.96R_\odot$, and the signs of the flows are randomly chosen. The calibration coefficients are obtained by performing a linear fit of the background flow map SHCs and the resulting travel time map SHCs at fixed $\ell$, as in Equation \ref{eqn:ineq}. The uncertainty of each calibration coefficient is given by the standard error computed from the residuals of the SHC linear fit at each $\ell$.

The equation used by \citet{Proxauf2021} and \citet{Hanasoge2012} to obtain the convective power spectrum with the above corrections is
\rev{\begin{equation}\label{eqn:p_spec_conv_P}
 (v_{\ell,m}^\phi)^2 \approx \dfrac{ (\delta\tau_{\ell,m}^{(\phi)})^2}{C_\ell^2}\dfrac{1}{D^2}\dfrac{1}{(N/S)^2}  ,
\end{equation}}
where $N/S$ is the inverse of the SNR that is specific to the measurement procedure and integration time, and $D$ is the ratio of the natural depth range of flows relative to the depth range of the prescribed flows. We use the same ratio as \citet{Hanasoge2012}, given in reference to the mixing length at $r=0.96R_\odot$, of $D=9.18$. We diverge slightly from the original methodology here by using a different factor to correct for the proportion of the measurement corresponding to the true flow signal. For a variable with zero mean, in this case, the longitudinal flow as observed from an appropriate reference frame, the corresponding SHCs obey
\[
\sum_{\ell,m}A_{\ell,m}^2 = \sigma^2,
\]
where $\sigma^2$ is the variance of the variable. We model the variance of the travel time measurements using the same formalism as the previous two works, as a sum of a time-independent signal ($S$) and a noise ($N_t$) component, the contribution of which decreases with the increase of the measurement time, $T$, as $1/T$
\begin{equation}\label{eqn:SNR}
\sigma^2 = S^2 + \dfrac{N^2}{T} = S^2 + N_t^2.
\end{equation}

The proportion of the power that comes from the time-independent signal is then
\[
\dfrac{S^2}{\sigma^2} = \dfrac{\sigma^2-N_t^2}{\sigma^2}
\]
and we use
\begin{equation}\label{eqn:p_spec_conv}
     (v_{\ell,m}^\phi)^2 \approx \dfrac{ (\delta\tau_{\ell,m}^{(\phi)})^2}{C_\ell^2}\dfrac{1}{D^2}\dfrac{S^2}{\sigma^2}  
\end{equation}
to compute the convective power spectrum from the longitudinal flow travel times. In Section \ref{sec:cal_results}, we present the derivation of the appropriate signal fraction, the results of this calibration procedure, and the validation of the derived calibration coefficients using a realistic background convective profile.

Keeping in line with previous works, the convective power spectrum is initially computed in terms of energy per multiplet,
\[
E_\phi(\ell) = \dfrac{1}{4\pi}\sum_{m\ne 0}(v_{\ell,m}^\phi)^2.
\]
In the convention of \citet{Hanasoge2012}, and further expanded upon in Appendix C of \citet{Birch2024}, this is converted to units of km$^2$~s$^{-2}$ per linear wavenumber (km$^3$~s$^{-2}$) through a multiplication of $r/2$. Here, $r$ is the corresponding radius at which the spectrum is measured. Substituting Equation \ref{eqn:p_spec_conv} into the previous equation yields the power spectrum in terms of the SHCs of the travel time differences,
\rev{\begin{equation}\label{eqn:P_spec_final}
E_\phi(\ell) = \dfrac{0.96R_\odot}{2}\dfrac{1}{4\pi}\sum_{m\ne 0}(v_{\ell,m}^\phi)^2 =  \dfrac{0.96R_\odot}{8\pi}\sum_{m\ne 0} \dfrac{ (\delta\tau_{\ell,m}^{(\phi)})^2}{C_\ell^2}\dfrac{1}{D^2}\dfrac{S^2}{\sigma^2}  .
\end{equation}}

\subsection{Observational Data Analysis} \label{sec:obs_method}

To obtain the travel time deviations corresponding to the longitudinal flows, we will apply the procedures outlined in Sections \ref{sec:df_method} and \ref{sec:cal_method} to Doppler velocity data from the Helioseismic and Magnetic Imager (HMI). We keep the observational cadence of 45~s, though the images are downsampled to a resolution of 0.25 degrees per pixel. This choice of resolution gives 720 pixels in latitude, which corresponds to a maximum resolvable spatial scale of $\ell=359$, well beyond the observational limit at $r=0.96R_{\odot}$.

We desire to include as much of the Sun's subsurface flow field as possible in the spherical harmonic transform, though this is complicated by the field of view of HMI being limited to one hemisphere. Limb foreshortening further complicates our measurements by decreasing the spatial extent of usable Doppler data to 70--80 heliographic degrees from disc center. We attempt to overcome this first limitation by constructing pseudo-synoptic travel time maps, while the second limitation can, in principle, be accounted for with a geometric correction.

First, we divide the full sphere into 24 segments in longitude, which results in 15$^\circ$, or 60 pixels, wide sections. The corresponding observation time, $T$, for each section is $1/24^\text{th}$ the Carrington rotation rate (27.275 days) or 1.136 days. The Dopplergrams for each of these sections are tracked at the Carrington rotation rate and centered at the heliographic longitude corresponding to the central meridian at half the observation duration. These tracked Dopplergrams are actually of size $720\times720$ pixels (one hemisphere) to better facilitate a spherical harmonic transform needed to apply the phase speed filter defined in Section \ref{sec:df_method}. The 60-pixel wide section is then extracted after the time-distance procedure is applied to the hemispheric maps. Furthermore, the section is cut beyond $\pm$65$^\circ$ where the travel time measurements become unacceptably noisy from limb foreshortening. This corresponds to a total reduction in power proportional to the missing area, where the integrated area is
\[
\int_\frac{5\pi}{36}^\frac{31\pi}{36}\int_0^{2\pi}\sin\theta d\phi d\theta = 3.625\pi \hspace{0.5em} \text{sr}.
\]
The computed power spectrum must then be corrected by a factor of $3.625\pi/4\pi=0.90625$.

\section{Results} \label{sec:results}

\subsection{Calibration and Validation} \label{sec:cal_results}

\begin{figure}
    \centering
    \includegraphics[width=0.8\linewidth]{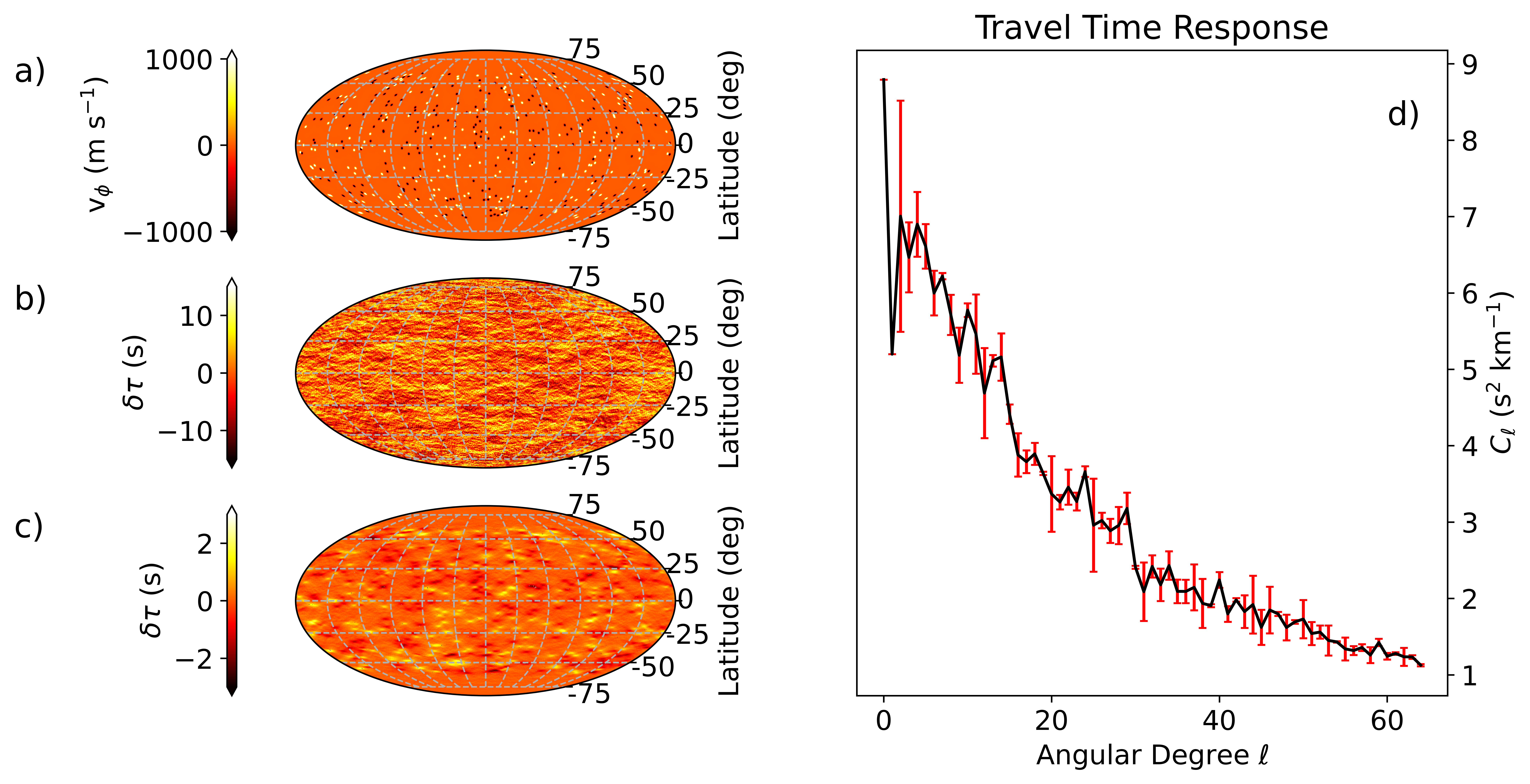}
    \caption{Results of the calibration procedure. Panel \textit{a} shows the input flow map with the 500 randomly placed "delta" flows. Panel \textit{b} shows the travel time deviations measured from the GALE simulation with the background flow map shown in panel \textit{a}. Panel \textit{c} shows the "clean" travel time deviations, after the contribution from the random sources has been removed. Panel \textit{d} shows the travel time response based on a linear fit of the SHCs from panels \textit{a} and \textit{c}; by definition, the error for the fit at $\ell=0$ and $\ell=1$ must be zero.}
    \label{fig:cal}
\end{figure}

We initiate a GALE simulation using the "delta" flows outlined in Section \ref{sec:cal_method} and we randomly distribute them within latitudes of $\pm65$$^\circ$ to better reflect the latitudinal extent of our solar observations. The input flow map is shown in Figure \ref{fig:cal}\textit{a}, and we note that the Gaussian shape of the flows is difficult to see at the global scale. The total simulation duration is 34 hours of solar time, though we keep only the final 24 hours to allow the high-$\ell$ \textit{p}-modes sufficient time to complete at least one round-trip. Discarding the first ten hours is also necessary since the surface velocity takes several hours to reach values consistent with solar convection, and the acoustic waves would not have yet fully interacted with the background flow, potentially leading to unrealistic noise in the time-distance measurements.

We then apply the time-distance procedure described in Section \ref{sec:df_method} to obtain the travel time signal associated with the longitudinal flows. While the amplitude of the flows is quite large, the small spatial extent leads to a relatively small contribution to the total travel time, and the resulting map is entirely dominated by noise (Figure \ref{fig:cal}\textit{b}). We can extract the flow signal from this map by subtracting the noise contribution. Since the only flows in the simulation are the ones prescribed, the only source of noise in the travel time maps must come from the random acoustic sources. We perform an additional simulation without background flows using the same sources as used for the "delta" flows and apply the time-distance procedure to obtain the travel time contribution from the random acoustic sources. This contribution is then subtracted from the initial time map to obtain the "clean" travel time deviations, i.e. those produced only by the background flow (Figure \ref{fig:cal}\textit{c}).

The SHCs for the input flow map and the resulting clean travel times are then computed to obtain $v_{\ell,m}$ and $\delta\tau_{\ell,m}$, respectively. The two coefficients are linearly related at each $\ell$ by the calibration coefficients $C_\ell$ described in Section \ref{sec:cal_method}, and a linear fit is performed to obtain these coefficients. We discard the $m=0$ component for $\ell>0$ as these correspond to differential rotation and do not contribute to the true convective power. The coefficients at $\ell=0$ and $\ell=1$ are derived slightly differently from those at other $\ell$. Since there is only one SHC at $\ell=0$, $C_0$ is given by dividing $\delta\tau_{0,0}$ by $v_{0,0}$, and so this coefficient must have zero error. We will admittedly not use $C_0$ since it has only the $m=0$ component, and we include it here only for comparative purposes. Similarly for $C_1$, there will only be two SHCs after excluding the $m=0$ component, so the error here must also be zero. We show the resulting calibration coefficients in Figure \ref{fig:cal}\textit{d} as a travel time response in units of s per km s$^{-1}$ of flow.

Our next step is to determine the proportion of the convective power that comes from the true flow signal, $S^2/\sigma^2$. We accomplish this using the Dopplergram data from halfway through CR 2184 (segment \#12) of duration 1.136 days and applying the time-distance procedure to increasingly longer segments of the data, ranging from 0.2 to 1 full durations or 0.227~days to 1.136~days. The travel time deviations for the 0.227~day and 0.341~day maps are, of course, extremely noisy, but some of these deviations take on unphysical values based on a failure of the Gabor wavelet fit. We therefore need to threshold the travel time maps to exclude these values, and this threshold must be carefully selected as it affects the derived variance. It is not clear \textit{a priori} what value to choose, and we select the threshold by identifying the value that minimizes the sum of the relative errors of the fit to Equation \ref{eqn:SNR}, the signal estimation $S$, and the noise estimation $N$. The function to be minimized is
\[
L(\delta\tau_\text{max}) = \dfrac{\sqrt{(\hat{\sigma}^2-\sigma^2)^2}}{\sigma^2} + \dfrac{\epsilon(S)}{S} + \dfrac{\epsilon(N)}{N},
\]
where $\sigma^2$ is the variance of the data, $\hat{\sigma}^2$ is the variance computed by fit, $\epsilon(S)$ is the uncertainty in $S$, and $\epsilon(N)$ is the uncertainty in $N$.

\begin{figure}[ht!]
    \centering
    \includegraphics[width=0.8\linewidth]{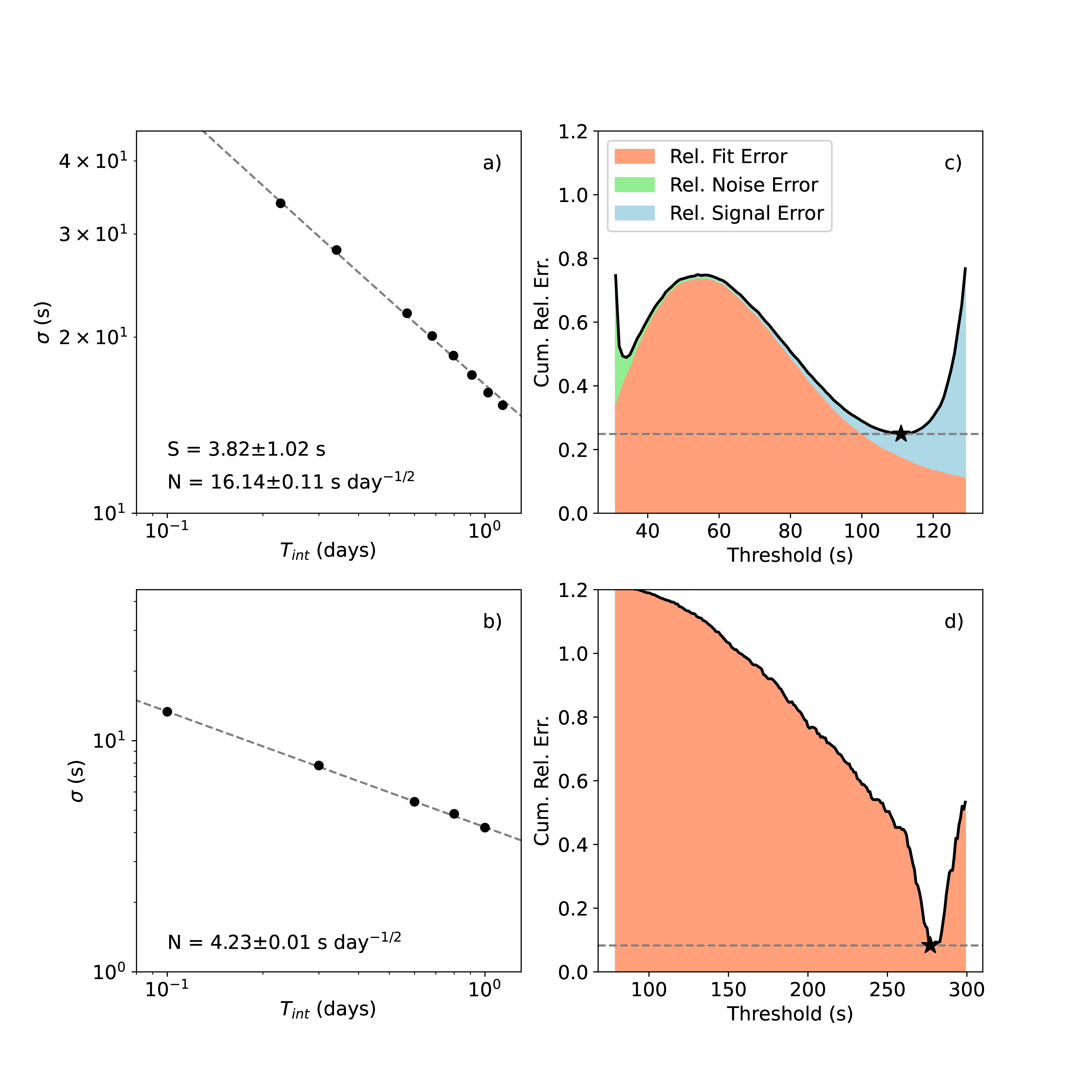}
    \caption{Panel \textit{a} shows the fit (dashed line) of the travel time variance of CR 2184 segment \#12 to the time-dependent noise and time-independent signal model. The threshold applied to the travel time maps (\textit{black star}) is chosen as the one that minimizes the cumulative relative error (panel \textit{c}) between the residuals (\textit{red}), signal uncertainty (\textit{blue}), and noise uncertainty (\textit{green}). Panels \textit{c} and \textit{d} show the same quantities for the noise-only model of the empty GALE simulation.}
    \label{fig:SN}
\end{figure}

The result of this minimization is shown in Figures \ref{fig:SN}\textit{a} and \ref{fig:SN}\textit{c} where the optimal threshold is $\delta\tau_\text{max}=111$~s that produces an estimated signal $S=3.82\pm1.02$~s and estimated noise $N=16.14\pm0.11$~s~day$^{-1/2}$ ($N_t=15.14\pm10$~s at 1.136~days); the corresponding correction factor is $S^2/\sigma^2=0.059\pm0.016$. We apply a similar procedure to the GALE simulations, with the added benefit that we can remove the time-independent signal by performing a simulation with no background flow for a more precise determination of the noise contribution. Figures \ref{fig:SN}\textit{b} and \ref{fig:SN}\textit{d} show the results of the previous procedure on the empty GALE simulation with an optimal threshold of $\delta\tau_\text{max}=277$~s and estimated noise $N=4.23\pm0.01$~s~day$^{-1/2}$. We discuss the derived thresholds and noise components in Section \ref{sec:dispcon}.

\begin{figure}
    \centering
    \includegraphics[width=0.8\linewidth]{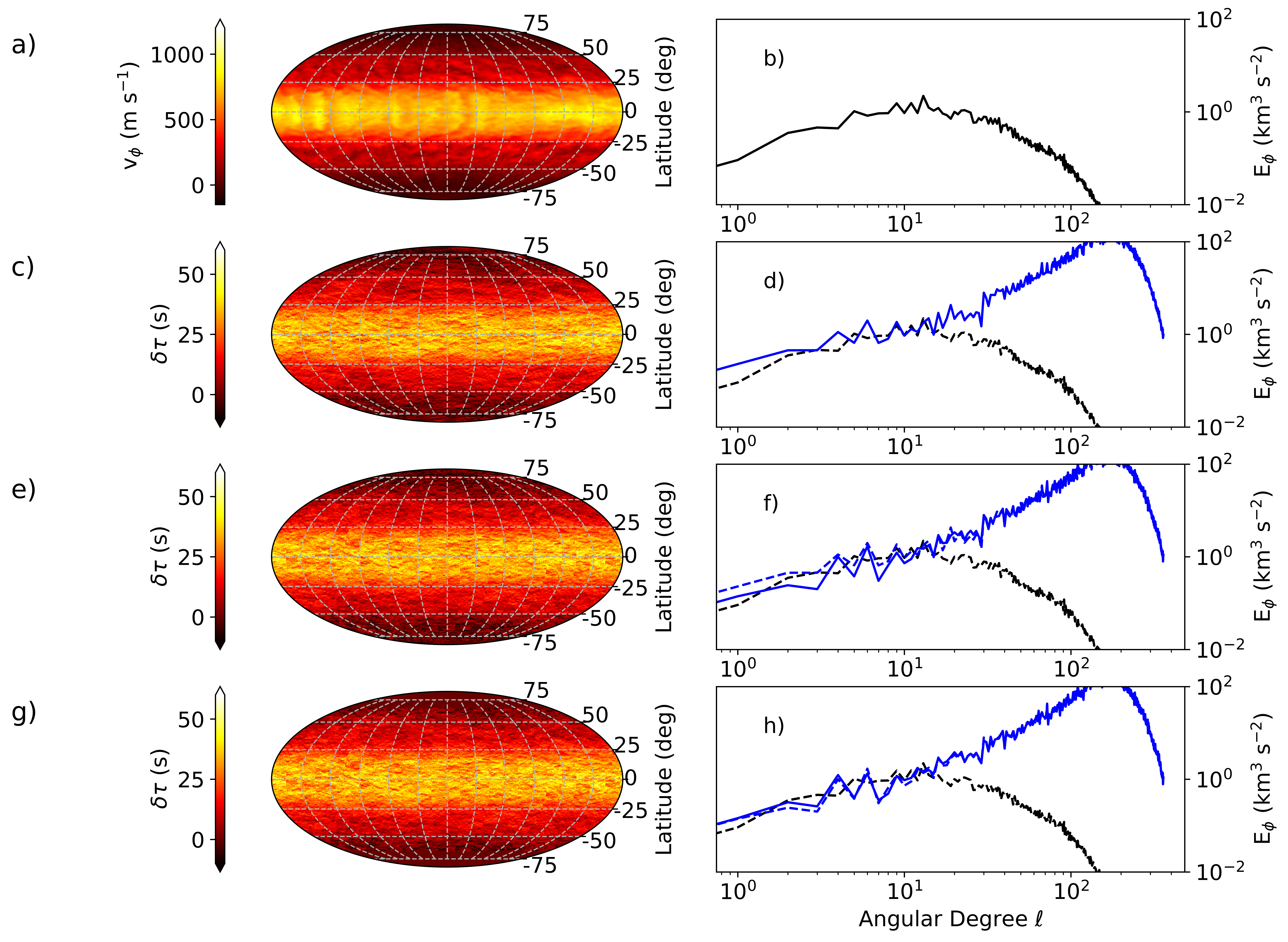}
    \caption{Effects of data reduction on the computed power spectrum. Panels \textit{a} and \textit{b} correspond to the input background flow field. Panels \textit{c} and \textit{d} correspond to the travel time measurements and derived power spectrum, respectively. In the third row, we replicate the construction of a pseudo-synoptic map from two independent simulations, with the power spectrum from panel \textit{d} shown with the blue dashed line. In the fourth row, the pseudo-synoptic map is truncated at the poles as in the solar data; the blue dashed line in panel \textit{h} corresponds to the power spectrum from panel \textit{f}.}
    \label{fig:val}
\end{figure}

We now assess the accuracy of the derived convective power spectrum using a GALE simulation initialized with a snapshot of the longitudinal flow from an EULAG simulation \citep{Smolarkiewicz2013}. The specific simulation, titled \texttt{R2x24}, was performed by \citet{Guerrero2022} with an initial solid body rotation period of 24~days, which eventually develops solar-like differential rotation. Since the EULAG simulation extends only up to $r=0.963\text{R}_{\odot}$, we artificially stretch the radial dependence of the flows up to R$_\odot$. This ensures that flows are populated at all points within the convection zone similar to within the Sun, which should reproduce similar noise characteristics in the travel time measurements. In our simulations, the flows at $r=0.96\text{R}_{\odot}$ correspond to those at $r=0.931\text{R}_\odot$ in the EULAG simulations. The input flow map from this simulation is shown in Figure \ref{fig:val}\textit{a}, and the corresponding convective power spectrum is shown in Figure \ref{fig:val}\textit{b}.

The pseudo-synoptic map is constructed from the travel time maps of two independent simulations initialized with the same background flow profile, where each individual segment of the maps is alternately sourced. The convective power in the pseudo-synoptic map (Figure \ref{fig:val}\textit{f} is not significantly different from that of the unaltered map (Figure \ref{fig:val}\textit{d}), though there is a slight systematic decrease below $\ell=3$ and for odd $\ell$. Since we do not expect the overall power to change moving from the unaltered map to pseudo-synoptic one, we do not apply any additional correction factor aside from those prescribed by Equation \ref{eqn:p_spec_conv_P}. We next substitute zeros for the measured travel times beyond $\pm65$$^\circ$ latitude (Figure \ref{fig:val}\textit{g}), as we will follow the same procedure for the solar data. We need to correct for the corresponding missing difference in power, dividing by a factor of $0.90625$, and the derived power spectrum is shown in Figure \ref{fig:val}\textit{h}. There is again very little difference in the convective power compared to the previous step in the data reduction, though the lack of power below $\ell=3$ is slightly ameliorated. Overall, we find that the power is reasonably well-captured by the deep-focus procedure up to $\ell=15$--$20$, but begins to diverge from the true spectrum beyond this range.

\subsection{\rev{Time-dependence of} the Solar Convective Power Spectrum} \label{sec:obs_results}

We finally apply the time-distance procedure and calibration coefficients to the Dopplergram data \rev{from Solar Cycle 24}. \rev{Beginning with CR 2101 and ending with CR 2227, the Dopplergram data are obtained for every 5$^{\text{th}}$ Carrington rotation and, when possible, the preceding Carrington rotation. The spectra from these two rotations are averaged to provide the power shown for each $\ell$ throughout the Solar Cycle in Figure \ref{fig:all_pow}. We use the monthly mean sunspot number (SSN), reported using the International Sunspot Number V2.0 standard published by the Royal Observatory of Belgium \citep{SSN}, to compare the evolution of the convective power with the progression of the Solar Cycle. Note that in Figure \ref{fig:all_pow}, we examine the time-dependence of the convective power only up to angular degree $\ell=15$, as it is likely that the spectrum is noise-dominated at smaller spatial scales.}

\begin{figure}
    \centering
    \includegraphics[width=\linewidth]{Figures/all_pow_tdep.png}
    \caption{\rev{The convective power at each $\ell$ is shown throughout Solar Cycle 24, with the corresponding angular degree labeled above each panel. Note that the scales for each panel are different in order to highlight the time-dependence. The individual measurements are shown as symbols with corresponding error bars, the black line shows the 13-month rolling average, and the gray line shows the monthly sunspot number. The black diamonds indicate an average of the central and preceding Carrington rotations, while the orange symbols denote a single-rotation measurement. Orange stars indicate that only the central Carrington rotation was available, while orange diamonds indicate that only the preceding Carrington rotation was available.}}
    \label{fig:all_pow}
\end{figure}

\rev{Overall, we find that the convective power does indeed show some variation as the Solar Cycle progresses, though for most spatial scales, this variation is within the error bounds and any dependence on the SSN is not particularly coherent. There is a general tendency for the power to increase in the first half of the Solar Cycle's active phase and a majority of the spatial scales have relatively low power during the second peak of the Cycle between early-2014 and late-2015. Of particular note, though, are the magnitudes of the convective power at $\ell=3$ and $\ell=5$ that show relatively coherent behavior throughout the Solar Cycle. In other words, the variation of the convective power at these scales is small from one Carrington rotation to the next, potentially indicative of some physical (signal-dominated) dependence on the phase of the Solar Cycle.}

\rev{We investigate this dependence in more detail with a correlation analysis using the time- and $\ell$-dependent power and the mean monthly SSN.  We compute the cross-correlation in the normalized sense, such that the correlation coefficient takes values between -1 (perfectly anti-correlated) and +1 (perfectly correlated). Specifically, we use
\[
C(\tau) = \sum_t \dfrac{\left(\text{SSN}(t+\tau)-\overline{\text{SSN}}\right)\left(E_\phi(t)-\overline{E_\phi}\right)}{\sigma(\text{SSN})\sigma(E_\phi)}, 
\]
where the overbar on each quantity denotes a time-average and $\sigma(x)$ refers to the standard deviation of quantity $x$. We compute both a forward and backwards cross-correlation such that a peak offset $\tau$ months indicates that the peak correlation occurs when the time-dependent convective power $E_\phi$ is shifted forwards by $\tau$ months; negative time lags correspond to a backwards shift of the time-dependent convective power. For this analysis, both the monthly sunspot number and the convective power have been interpolated to be uniformly spaced in time with cadence 1 month, which we define as $30.4$ days.}

\begin{figure}
    \centering
    \includegraphics[width=0.8\linewidth]{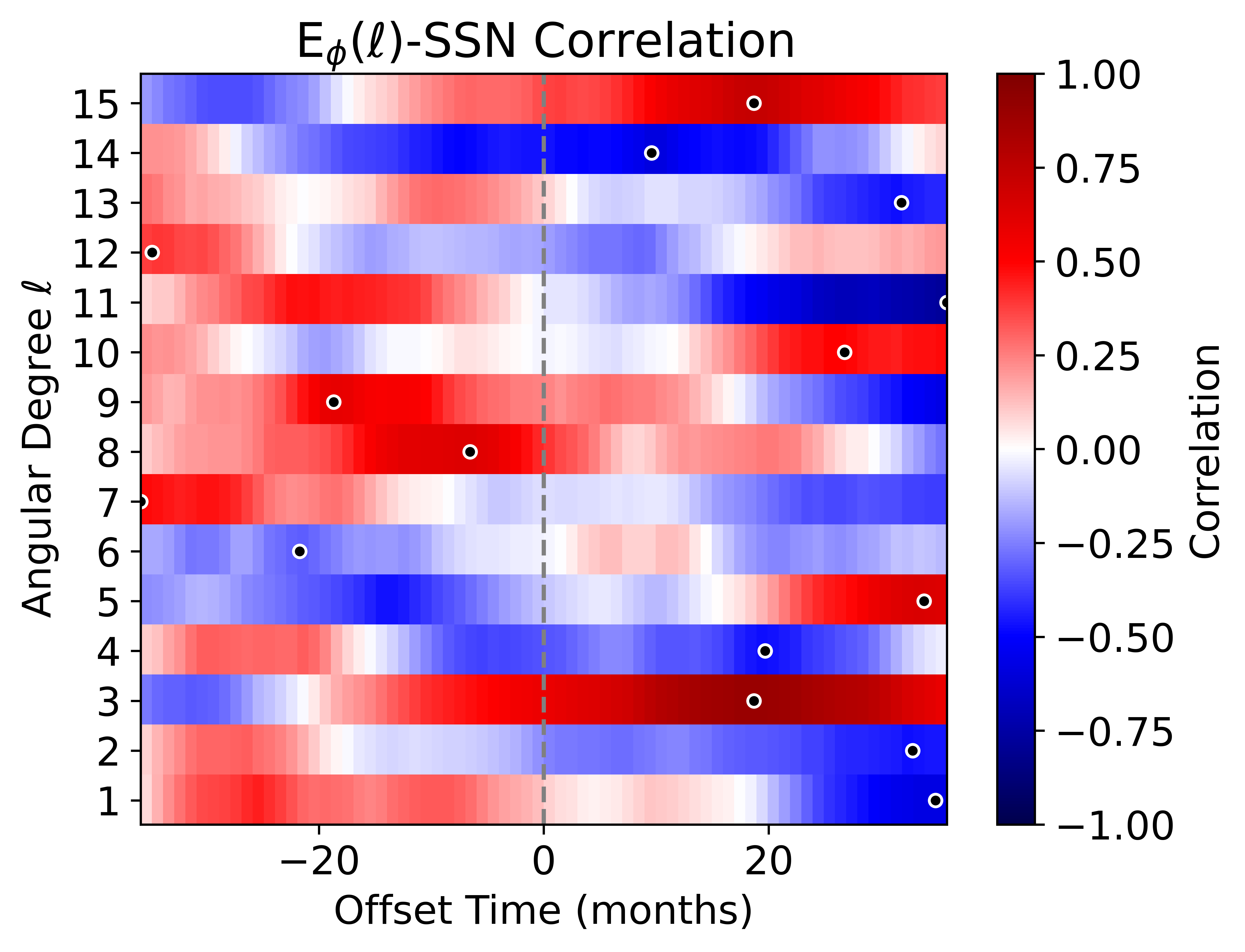}
    \caption{\rev{Correlations, as a function of time lag (x-axis) and spatial scale (y-axis), between the monthly mean SSN and the convective power. Red colors correspond to correlations, while blue colors correspond to an anti-correlation. The greatest absolute correlation for each angular degree is marked by a black and white circle.}}
    \label{fig:corr}
\end{figure}

\rev{These correlation functions are shown in Figure \ref{fig:corr}. While we observe some strong correlations, particularly for $\ell=3$ ($C=0.85$ at $\tau=17.7$\,months) and $\ell=15$ ($C=0.78$ at $\tau=19.7$\,months), the width of these correlation peaks makes it difficult to assess how reliable these signatures are. A visual inspection of the power at $\ell=15$ in Figure \ref{fig:all_pow} does hint at some kind of cycle dependence, but the time between peaks in the power is not quite as long as for $\ell=3$ and does not appear to span the full length of the Solar Cycle. There are not any particularly strong anti-correlations between the convective power and the SSN, though it is worth mentioning that $\ell=11$ shows the strongest anti-correlation ($C=-0.82)$ at a time lag of $\tau=35$\,months. However, visually inspecting the time-dependence of the convective power at this spatial scale does not reveal any immediately noticeable trends.}

\begin{figure}
    \centering
    \includegraphics[width=0.8\linewidth]{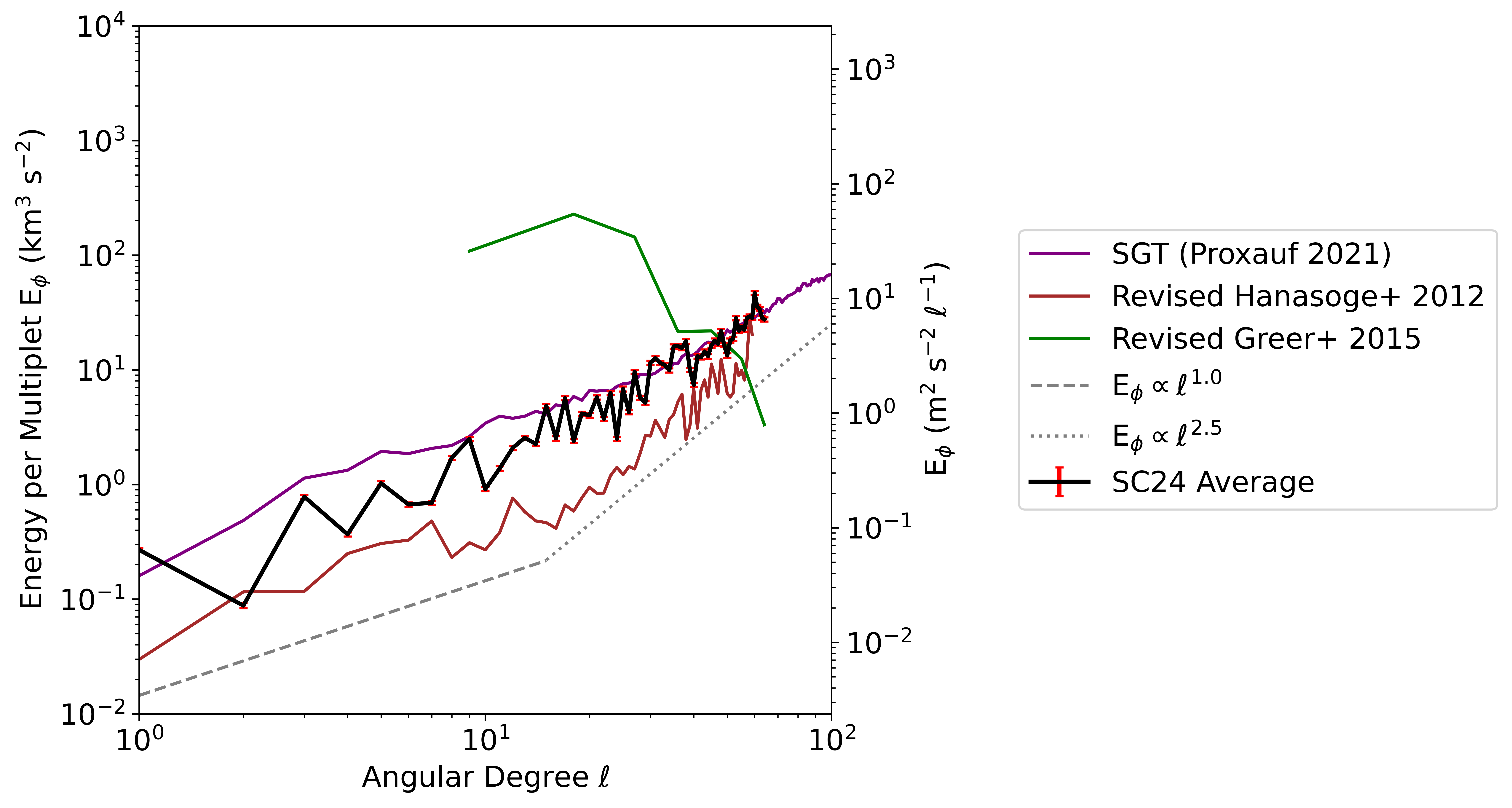}
    \caption{\rev{Our convective power spectrum for Solar Cycle 24 obtained from averaging across the spectra from each analyzed Carrington rotation (black). For comparison, the original time-distance spectrum \citep{Hanasoge2012} and ring-diagram spectrum \cite{Greer2015} obtained at $0.96$R$_\odot$ are shown in maroon and green, respectively. The purple line shows the convective power at R$_\odot$ derived from surface granulation tracking \citep{Proxauf2021}.}}
    \label{fig:av_pow}
\end{figure}

\rev{Finally, we obtain an average convective power spectrum across the entire Solar Cycle by averaging across every analyzed Carrington rotation (N=47). This spectrum is presented in Figure \ref{fig:av_pow} with the error bars shown to scale. We find that the spectrum is qualitatively similar in slope to the original time-distance result \citep{Hanasoge2012}, increasing proportional to $\ell$ until a transition to the noise-dominated regime where the power increases as $\sim\ell^{2.5}$. We note that the dependence on $\ell$ of our spectrum in the noisy regime increases slightly less dramatically than the \citet{Hanasoge2012} spectrum ($E_\phi \propto \ell^{2.1}$), though this does not make the power in this range of angular degrees any more reliable. Still, it can be concluded from the validated portion of the spectrum that the power consistently increases with angular degree, aside from some minor variations. We do not observe a "turn-off" in the power spectra in this regime, and it is likely that the convective power peaks at some angular degree greater than $\ell=15$, corresponding to spatial scales smaller than 291\,Mm.}

\section{Discussion \& Conclusion} \label{sec:dispcon}

Our validation procedure, outlined in Figure \ref{fig:val}, gives us reasonable confidence in our computed solar convective power spectrum at least up to $\ell\approx 15$. The power does diverge from the ground truth between $\ell=15$ and $\ell=30$, and it is difficult to assign a specific cut-off scale given the measurement uncertainty and $\ell$-to-$\ell$ variation in the power. Looking at how the convective power grows in our measurements and those of \citet{Hanasoge2012}, we see similar characteristic growth rates in both the reliable and unreliable ranges. In general, the power in the reliable range grows like $E_\phi\propto\ell^1$ and in the unreliable range as $E_\phi\propto\ell^{2.5}$ (the dashed and dotted curves in Figure \ref{fig:av_pow}, respectively). The growth rate is slightly larger in our measurements than in \citet{Hanasoge2012} ($E_\phi\propto\ell^{1.06}$ vs $E_\phi\propto\ell^{0.97}$, though the growth rate is similar in the unreliable---presumably noise-dominated---range.

We can compare our power spectrum to measurements made using other techniques, though the only work that measures the spectrum near $r=0.96R_\odot$ is the ring-diagram analysis performed by \citet{Greer2015} and refined by \citet{Proxauf2021}; this is shown as the green curve in Figure \ref{fig:av_pow}. We now know that our convective power is unreliable beyond $\ell=30$, and perhaps even lower, so direct comparison of the two spectra beyond this range is unreasonable. We therefore note that the disagreement beyond $\ell=30$ cannot be fairly evaluated. We focus instead on the overlapping range of measurements between $\ell=10$ and $\ell=20$, where there is \rev{still a stark difference} in the power computed from the time-distance method and the ring-diagram method. \rev{Though the overlapping region is towards the less-reliable end of our measurements, a $100\times$ increase in the convective power between $\ell=10$ and $\ell=15$ to align the reliable range of our spectrum with the ring-diagram spectrum is not physically well-justified.}

It is also worth discussing what we might expect of the convective power spectrum for characteristic scales smaller than ($\ell$ greater than) our reliable threshold, particularly in the context of supergranulation that appears as a peak near $\ell_\text{peak}=150$ in surface measurements \citep{Roudier2012,Proxauf2021}. The placement of this peak varies as a function of depth, and it has been shown in power spectra of $\text{div}(\mathbf{v})$ that $\ell_\text{peak}$ decreases slightly with depth \citep[see Figure 7 of][]{Getling2022}. Additionally, the peak broadens with depth, which implies a superposition of characteristic flows as opposed to the single peak predicted by mixing length theory (MLT) corresponding to the mixing length $\lambda=\alpha_\text{MLT}H_P$, where $\alpha_\text{MLT}$ is the mixing length parameter and $H_P$ is the pressure scale height. \rev{In MLT, the mixing length $\lambda$ is the characteristic distance a parcel of fluid rises or sinks before equilibrating with the local fluid and is related to the (essentially unfixed) mixing length parameter $\alpha_\text{MLT}$, which must be constrained by stellar evolutionary models. The mixing length parameter depends on a variety of factors, including initial helium and heavy element abundance, and for the Sun it is estimated that $\alpha_\text{MLT}\approx 1.8 \text{--} 1.9$ \citep{Joyce2018}. The mixing length at $r=0.96R_\odot$, computed with pressure scale height estimated from the Standard Solar Model, is $\lambda\approx 18.5$~\,Mm which corresponds to angular degree $\ell\approx 236$. This scale is slightly smaller than the photospheric supergranular scale, and} it is tempting to extrapolate the $\ell$-dependence of our measurements to these scales in order to estimate the strength of such flows in the interior. However, the unreliability of our measurements in this range coupled with the extremely large amplitudes implied by this extrapolation leaves little room for a meaningful analysis.

\rev{Furthermore, a physical interpretation of potential correlation between the solar cycle and magnitude of the convective power at $r=0.96$R$_\odot$ is not exactly clear. At the photosphere, at least, it is well-known that the presence of strong magnetic fields inhibits local convection and granules tend to become smaller at moderate field strengths \citep{Nordlund2009}. One might expect similar behavior for subsurface convection in the presence of magnetic fields, resulting in a suppression of low-$\ell$ convective power and potential enhancement at higher angular degrees. Our most coherent angular degrees (in time) at $\ell=3, 5$ and $15$ paint a more complicated picture. The convective power at all three angular degrees show an increase in the first half of the Solar Cycle's active phase, and it is only near the end of solar maximum that the power is suppressed. Additionally, only $\ell=5$ shows an absolute minimum during this second peak of solar maximum, while the power at $\ell=3$ decreases throughout the active phase before increasing again in the declining phase. The convective power at $\ell=15$ changes in a qualitatively similar fashion to the power at $\ell=3$, so the change in convective scales with increasing magnetic field strength may occur at finer spatial scales that are unresolved by our measurements.}

To conclude, we have used the deep-focus time-distance methodology as outlined by \citet{Hanasoge2012} to obtain new measurements of the solar convective power spectrum. We validated this methodology using simulations as ground truth and found that the derived power spectrum diverges from the true spectrum beyond $\ell=15$--$30$. \rev{In the reliable range, the average of convective power over Solar Cycle 24 is qualitatively similar to the original result, showing similar growth rates in the reliable and unreliable regimes. While the amplitude of this spectrum is slightly higher than the original result by about half an order of magnitude, this increase is insufficient to resolve the Convective Conundrum that persists between simulations of global convection and observations.}

Moving forward, we plan on applying the same validation procedure employed here to the ring-diagram methodology used by \citet{Greer2015}. This will allow us to make a direct comparison between the two spectra, which we hope will highlight limitations and inconsistencies that may reveal a more constrained solar power spectrum.

\begin{acknowledgements}
    This work was partially supported by NASA grants {80NSSC19K0268,  80NSSC19K1436, 80NSSC20K1320, and 80NSSC23K0097}. Resources supporting this work were provided by the NASA High-End Computing (HEC) Program through the NASA Advanced Supercomputing (NAS) Division at Ames Research Center. We thank Quentin Noraz and collaborators, as well as Hideyuki Hotta and collaborators, for sharing the convective power spectrum data from recent simulations.
\end{acknowledgements}

\bibliography{CC_DF}{}
\bibliographystyle{aasjournal}

\end{document}